\documentclass[prb,twocolumn]{revtex4-1}
\usepackage{amsfonts}
\usepackage{amsmath}
\usepackage{amssymb}
\usepackage{graphicx}
\begin{document}

\title{Tunneling characteristic of a chain of Majorana bound states}

\author{Karsten Flensberg}
\affiliation{Department of Physics, Harvard University,
Cambridge, Massachusetts 02138, USA\\Niels Bohr Institute,
University of Copenhagen, Universitetsparken 5, DK-2100
Copenhagen, Denmark}

\begin{abstract}
We consider theoretically tunneling characteristic of a
junction between a normal metal and a chain of coupled Majorana
bound states generated at crossings between topological and
non-topological superconducting sections, as a result of, for
example, disorder in nanowires. While an isolated Majorana
state supports a resonant Andreev process, yielding a zero bias
differential conductance peak of  height $2e^2/h$, the
situation with more coupled Majorana states is distinctively
different with both zeros and $2e^2/h$ peaks in the
differential conductance. We derive a general expression for
the current between a normal metal and a network of coupled
Majorana bound states and describe the differential conductance
spectra for a generic set of situations, including regular,
disordered, and infinite chains of bound states.
\end{abstract}
\date{\today} \maketitle

Topological materials are of large current interest, in part
because of their potential for topological quantum computing
and their interesting non-Abelian
quasiparticles.\cite{Nayak2008} One variant of this is
topological superconductors where the low energy quasiparticles
in addition are Majorana Fermions.\cite{Kitaev2001,Fu2008}
Currently, there is an active search for materials that can
host such particles, either in certain $p$-wave
superconductors, or semiconductors with proximity induced
superconductivity and strong spin-orbit
coupling.\cite{Oreg2010,alicea2010,Lutchyn2010,Sau2010} Because
it takes two Majorana Fermions to form a usual Dirac Fermion
that can couple to other degrees of freedom, detection of the
\textit{state} of the Majorana fermion system requires
non-local measurements or
interferometry.\cite{Fu2009,Akhmerov2009,Fu2010} In contrast, a
local tunnel current, being independent of the parity of the
topological superconductor, does not reveal information about
the state of the Majorana fermions. Nevertheless, a tunneling
probe could detect the \textit{presence} of a Majorana bound
state (MBS) \cite{Law2009,Linder2010,Sau2010} and the detection
of Majorana bound states is the first major challenge in this
field. Tunneling contact to an isolated Majorana state give
rise to a resonant Andreev process that gives a zero bias
conductance peak of $2e^2/h$.\cite{Law2009} With two coupled
Majorana states cross correlations of the current into each
could also show their existence and non-local
character.\cite{Bolech2007,Nilsson2008}

However, because of material difficulties, isolated Majorana
states might be rare. Rather it is to be expected that density
fluctuations will generate a random configuration of
topological/non-topological boundaries, at which Majorana
states will be located. For strong disorder the distances
between these states are sufficiently short for the MBS to
overlap and therefore it is important to understand how a
network of coupled Majorana fermions maps onto the tunneling
characteristic. This problem was recently considered in
Ref.~\onlinecite{Shivamoggi2010} in the weak coupling regime,
using a renormalization group to reduce a chain to a sum of
single Majorana pairs on a logarithmic energy scale.

In this paper, a theory for tunneling between a metallic probe
and a collection of coupled Majorana states in the strong
coupling regime is developed, and experimentally relevant
situations are addressed. This is done in the limit where the
voltage $eV$, the tunneling broadenings $\Gamma$, and the
hopping matrix elements between any two MBS all are much
smaller than the superconducting energy gap $\Delta$. The
regime of large $\Delta$ is well suited for characterization
and detection of the MBS, because in absence of Majorana states
the Andreev conductance is of the order\cite{BTK}
$(e^2/h)(\Gamma/\Delta)^2$ and thus much smaller than the
resonant Andreev current carried by the Majorana states.
Examples for different number and configurations of Majorana
states is given and the case of a uniform infinite chain is
solved exactly. Finally, disordered Majorana chains are
addressed. Disorder is introduced as random nearest neighbor
couplings and it is show to reduce to a finite chain, truncated
by the first weak link (quantified below) in the chain.
\begin{figure}[ptb] \centering
\includegraphics[width=.5\textwidth]{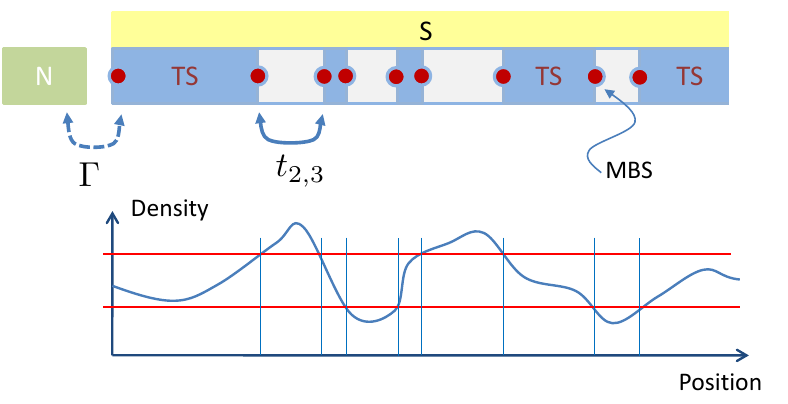}
\caption{ (Color online) Illustration of a semiconductor with induced
superconducting order parameter from an adjacent superconductor (S).
Spatial variations in density or
superconducting order parameter create crossings between
topological (TS) and non-topological segments when the density crosses
the critical density (vertical (red) lines). Majorana bound
states (MBS) (circles) are located at each crossing point and the distances
between them determine the coupling matrix elements $t_{ij}$ of the resulting
Majorana network. A tunneling contact (N) probes the network by
tunneling into the end Majorana mode, with tunneling density of states $\Gamma$.} \label{fig:model}
\end{figure}

The borders  of the topological superconductor segments give
rise Majorana bound states. These states are zero energy
solutions to the Boguliubov-de Gennes equations for the
geometry in question. The general form of a MBS is
\begin{equation}
\gamma_{i}^{{}}=\sum_{\sigma}\int dx~\left(  f_{\sigma,i}^{{}}(x)\Psi_{\sigma
}^{{}}(x)+f_{\sigma,i}^{\ast}(x)\Psi_{\sigma}^{\dagger}(x)\right)  .
\end{equation}
The Majorana Fermion has the properties that
$\gamma_{i}^{{}}=\gamma _{i}^{\dagger}$ and $\gamma_{i}^{2}=1$.
The superconductor Hamiltonian, describing the coupled Majorana
state network, is
\begin{equation}
H_{S}=\frac{i}{2}\sum_{ij}t_{ij}^{{}}\gamma_{i}^{{}}\gamma_{j}^{{}},
\end{equation}

The tunnel Hamiltonian between the normal metal and the
superconductor is
\[
H_{T}=\sum_{k\sigma}\int dx~\left(  t_{k}^{\ast}(x)c_{k\sigma}^{\dagger}
\Psi_{\sigma}^{{}}(x)+\mbox{h.c.}\right),
\]
where $c_{k,\sigma}$ are lead-electron annihilation operators
and $\Psi_\sigma(x)$ the superconductor electron-field
operator. As explained above, for $(eV,\Gamma)\ll\Delta$ the
Majorana states contribution to the current dominates. Using
Nambu representation,
$\boldsymbol{\Psi}=(\Psi_{\uparrow},\Psi_{\downarrow},\Psi_{\downarrow}^{\dagger},\Psi_{\uparrow}^{\dagger})$,
the projection of the field operator $\boldsymbol{\Psi}$ onto
the manifold of Majorana states is
$\boldsymbol{\Psi}(x)\approx\sum_{i}\gamma_i
(f_{\uparrow,i}^{{}}(x),f_{\downarrow,i}^{{}}(x),f_{\downarrow,i}^{\ast}(x),f_{\uparrow,i}^{\ast}(x))$,
which then leads to the effective tunnel Hamiltonian describing
the coupling between the lead and the Majorana states
\begin{equation}
H_{T}=\sum_{k\sigma i}(V_{k\sigma,i}^{\ast}c_{k\sigma}^{\dagger}-V_{k\sigma,i}^{{}}
c_{k\sigma}^{{}})\gamma_{i}, \label{Hteff1}
\end{equation}
where $ V_{k\sigma,i}=\int dx\,f_{\sigma,i}^{{}}(x)t_{k}(x)$.
The current operator is given by the rate of change of the
number of electrons in the normal lead
\begin{equation}
I  =-e\dot{N}=-ie[H_{T},N]/\hbar=\frac{2e}{\hbar}\operatorname{Re}\sum_{k\sigma i}\left(  V_{k\sigma,i}^{\ast
}G_{i,k\sigma}^{<}(0)\right)  ,
\end{equation}
where the lesser Green's function combining $k\sigma$ and $i$
is defined as $G_{i,k\sigma}^{<}(t)=i\left\langle
c_{k\sigma}^{\dagger }\gamma_{i}(t)\right\rangle$, which is
written as
\begin{equation}
G_{i,k\sigma}^{<}(t)=\sum_{j}\left[  G_{ij}V_{k\sigma,j}^{{}}G_{k\sigma}
^{(0)}\right]  ^{<},
\end{equation}
where $G_{ij}(\tau,\tau^{\prime})=-i\left\langle
T(\gamma_{i}(\tau)\gamma _{j}(\tau^{\prime})\right\rangle $ is
the full Keldysh time-ordered Green's functions for the
Majorana operators, and $G_{k\sigma}^{(0)}(\tau,\tau
^{\prime})=-i\left\langle
T(c_{k\sigma}^{{}}(\tau)c_{k\sigma}^{\dagger}
(\tau^{\prime})\right\rangle _{0}$ is the unperturbed normal
lead Green's function. By choosing the chemical potential of
the superconductor as a reference, the general current formula
is derived to be (see Appendix)
\begin{equation}\label{Current}
I=\frac{e}{h}\int
d\omega~M(\omega)\left[f(-\omega+eV)-f(\omega-eV)\right],
\end{equation}
with $f$ being the Fermi-Dirac distribution and
\begin{equation}\label{Mdef}
M(\omega)=\mathrm{Tr}\left[\mathbf{G}^{R}(\omega)
\mathbf{\Gamma}^{\ast}(-\omega)\mathbf{G}^{A}(\omega)
\boldsymbol{\Gamma}(\omega)\right].
\end{equation}
Here the retarded Majorana Green's function is
\begin{equation}
\mathbf{G}_{\omega}^{R}=2\left(  \omega-2i\mathbf{t}+i\left(  \mathbf{\Gamma
}_{\omega}+\mathbf{\Gamma}_{-\omega}^{\ast}\right)  -\left(  \mathbf{\Lambda
}_{\omega}-\mathbf{\Lambda}_{-\omega}^{\ast}\right)  \right)  ^{-1},
\end{equation}
where $\mathbf{t}$ is an antisymmetric matrix, while the
Hermitian matrices $\mathbf{\Gamma}$ and $\mathbf{\Lambda}$ are
\begin{align}\label{Gammadef}
\Gamma_{ij}(\omega) &  =2\pi\sum_{k\sigma}V_{k\sigma,i}^{{}}V_{k\sigma,j}^{\ast}\delta\left(
\omega-\varepsilon_{k\sigma}\right)  ,\\
\Lambda_{ij}(\omega) &  =\mathcal{P}\!\int\frac{d\omega^{\prime}}{2\pi}
\frac{\Gamma_{ij}(\omega^{\prime})}{\omega-\omega^{\prime}}.
\end{align}
If the coupling matrix respects particle-hole symmetry,
$\mathbf{\Gamma }(\omega)=\mathbf{\Gamma}^{\ast}(-\omega)$, the
current is antisymmetric $I(V)=-I(-V)$ (see Appendix for more details).

The expression (\ref{Current}) is a general finite temperature
expression for the current into a Majorana state network in
terms of matrices describing the coupling to the normal lead
and the Majorana network. The general formula is
straightforwardly extended to the case with more normal metal
contact connected to the network.\cite{multi}

If the Majorana bound states are separated in space by a
distance much longer than the normal metal Fermi wavelength,
off-diagonal terms of $\Sigma _{ij}^{R}$ will tend to average
out due to the fast variation of the phase of
$V_{k\sigma,i}^{{}}$. In this case, it is a good approximation
to set $\Gamma_{ij}^{R}(\omega)\approx
\delta_{ij}\Gamma_{ii}(\omega)$, which is assumed from here on.
Moreover, assuming a weak energy dependence $V_{k\sigma,i}$ and
hence a constant $\Gamma_{ii}$, so that $\Lambda_{ij}=0$,
(so-called wide-band limit), the differential conductance
reduces to
\begin{equation} \label{didvfinal}
\frac{dI}{dV}=\frac{2e^{2}}{h}\int d\omega~
\operatorname{Im}\sum_{i}\left[\Gamma_{ii}G_{ii}^{R}(eV)\right]
\left(\frac{df(\omega-eV)}{d\omega}\right),
\end{equation}
with
\begin{equation}\label{GR}
\mathbf{G}^{R}(\omega)=2\left[\omega-2i\mathbf{t}+i2\mathbf{\Gamma}\right]
^{-1},
\end{equation}
and $\mathbf{\Gamma}$ is a diagonal matrix.

\begin{figure}[ptb]
\centering
\includegraphics[width=.45\textwidth]{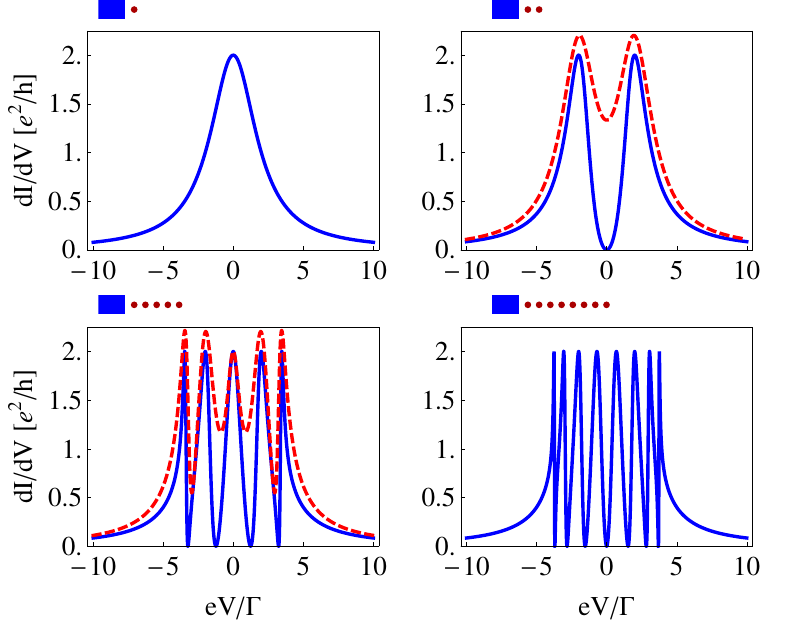} \caption{ (Color online) The conductance for
tunneling into the end of a chain with 1,2,5, and 8 Majorana states
coupled by $t=\Gamma$. A general feature, valid also for disordered arrays, is
that for even number of coupled Majorana modes, the conductance is zero at
zero voltage, whereas for an odd number it is given by $2e^{2}/h.$ Moreover,
the conductance has in general $n-1$ zeros, where $n$ is the number of
Majoranas in the cluster that couples to the probe.
The dashed (red) curves show the result when the two first Majorana states both
coupled to the lead, the second one with strength $\Gamma_{22}=\Gamma_{11}/2$.
}
\label{fig:gfew}
\end{figure}

For a single isolated Majorana state with tunnel broadening
coupling the Green's function is: $G_{ii}^{R}=2/(\omega
+i2\Gamma)$ and the zero temperature differential conductance
is easily obtained as
\begin{equation}
\frac{dI}{dV}=\frac{2e^{2}}{h}\frac{4\Gamma^{2}}{\left(  eV\right)  ^{2}
+4\Gamma^{2}},
\end{equation}
which confirms that the resonant Andreev tunneling with zero
bias conductance $G=2e^{2}/h$.\cite{Law2009} With two Majoranas
coupled by tunneling $t$ and only one of them coupled to the
lead the differential conductance is
\begin{equation}
\frac{dI}{dV}=\frac{2e^{2}}{h}\frac{(2eV\Gamma)^{2}}{\left(  (eV)^{2}
-4t^{2}\right)  ^{2}+(2eV\Gamma)^{2}},
\end{equation}
which has a dip at zero voltage and peaks at $eV=\pm2t$, where
the conductance again reaches $2e^{2}/h.$ In fact, a very
general statement holds for tunneling into the end of a chain,
namely that \textit{with an odd number of coupled Majorana
states the zero bias conductance is $2e^2/h$, and with an even
number the zero bias conductance is zero}. This can be shown by
the inversion in Eq.~\eqref{GR} setting $\omega=0$ and for an
arbitrary chain matrix $t_{ij}$. Moreover, \textit{for a
cluster with $n$ MBS the differential conductance versus bias
voltage has $n-1$ zeros and $n$ voltages where $dI/dV=2e^2/h$.}
If the normal metal electrode overlaps with more than one MBS
these conclusions change, as shown in Fig.~\ref{fig:gfew},
where the conductance for some examples is plotted.

\begin{figure}[ptb]
\centering
\includegraphics[width=.4\textwidth]{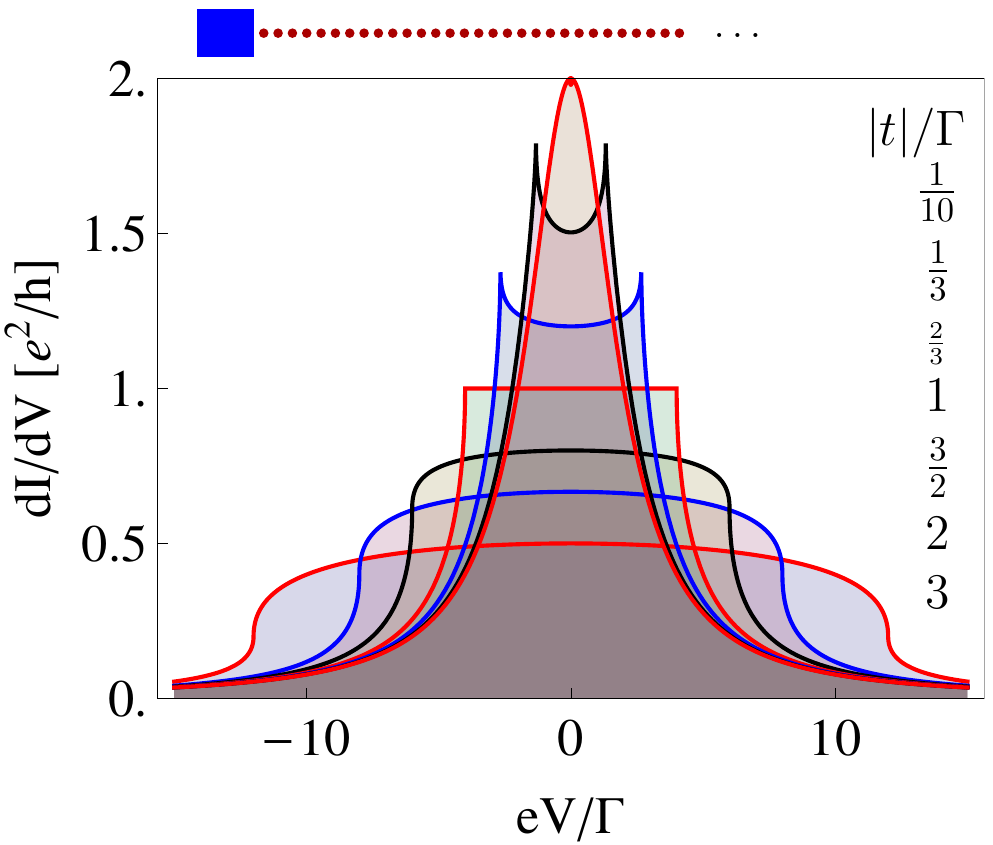} \caption{(Color online) The conductance for
tunneling into the end of an infinite Majorana chain with identical tunneling
couplings. The parameters $|t|/\Gamma$ ranges from 5 to $1/8$, from outside in.}
\label{fig:ginf}
\end{figure}
With many coupled MBS in the chain the conductance oscillates
between $2e^2/h$ and 0, as seen in Fig.~\ref{fig:gfew}. As the
number of sites in the chain is increased, the period of the
oscillations decreases. If the period is smaller than
temperature the conductance will average to a value between the
two extremes. The same occurs for an infinite homogeneous
chain, which is considered next.

With an infinite chain of Majorana states with nearest neighbor
couplings , $t_{ij}$, the Green's function for the first MBS is
$G_{11}^R=2g_{11}^{{}}$, where
\begin{equation}\label{g11}
g_{11}^{{}}=\frac{1}{(g_{11}^{0})^{-1}-4|t_{12}|^2\tilde{g}_{22}^{{}}},
\end{equation}
and where $\tilde{g}_{22}^{{}}$ is the Green's function for the
network starting with site 2, decoupled from site 1, and where
$(g_{11}^{0})^{-1}=\omega+2i\Gamma$. An illustrative example is
a homogeneous chain, i.e. with all couplings identical
$t_{ij}=t$. Then the Dyson equation for $\tilde{g}_{22}^{{}}$
is
\begin{equation}
\tilde{g}_{22}^{{}}=\frac{1}{\omega+i\eta}+\frac{4t^{2}}{\omega+i\eta}\tilde{g}_{33}^{{}
}\tilde{g}_{22}^{{}}.
\end{equation}
Since all connections are equal
$\tilde{g}_{22}^{{}}=\tilde{g}_{33}^{{}},$ and hence
$(4t^{2}/\omega)\tilde{g}_{22}^{2}-\tilde{g}_{22}^{{}}+1/\omega=0$,
which gives can be solved for $\tilde{g}_{22}$. Choosing the
correct branch cuts\cite{branchcuts} and setting this into the
Green's function \eqref{g11}, the differential conductance is
derived to be
\begin{equation}
\frac{dI}{dV}=\frac{2e^{2}}{h}\left\{
\begin{array}
[c]{cc}
\frac{4\Gamma\left(4\Gamma+\sqrt{(4t)^{2}-(eV)^{2}}\right)}{(eV)^{2}+\left(4\Gamma+\sqrt{\left(  4t\right)  ^{2}-(eV)^{2}}\right)
^{2}}, & |eV|<4|t|,\\
\frac{(4\Gamma)^2}{\left(  |eV|+\sqrt{(eV)^{2}-(4t)^{2}}\right)
^{2}+(4\Gamma)^{2}}, & |eV|>4|t|.
\end{array}
\right.
\end{equation}
This is an interesting expression with a line shape that
strongly depends on the ratio $t/\Gamma$, which is shown in
Fig.~\ref{fig:ginf}. Furthermore, for $eV=0$ it reduces to
\begin{equation}
\left.  \frac{dI}{dV}\right\vert _{V=0}=\frac{2e^{2}}{h}\frac{2\Gamma}
{|t|+2\Gamma},
\end{equation}
which shows that for small tunnel broadening compared to the
bandwidth of the chain the zero bias conductance is $2e^{2}/h$,
which was expected because it corresponds to tunneling into an
effectively isolated Majorana state.

As the last situation, which might also be the most
experimentally relevant, we now discuss different realizations
of long disordered chains, sampled for example by scanning the
average density and creating a different set of crossings of
the topological superconductor thresholds, as illustrated in
Fig.~\ref{fig:model}. One could in this way study the average
conductance of a network, given a distribution of nearest
neighbor MBS couplings $t_{ij}$. The tunneling coupling between
two neighboring MBS depends both on the distance between them
and the deviation from the critical value for the
topological/non-topological transition, with exponential
dependence on both, as was shown by Shivamoggi et
al.\cite{Shivamoggi2010} for a specific example. The
distribution of tunneling couplings is thus a complicated
convolution of amplitude fluctuations and the level-crossings
statistics,\cite{levelcrossing} in this case with two
crossings. The resulting distribution of tunneling couplings is
an interesting problem in itself. However, instead of pursuing
this line, we focus at the generic behavior expected for a
given configurations of the tunneling couplings in the chain.

As we learned for the infinite chain, different behaviors occur
depending on the ratio of $\Gamma$ to the tunneling couplings
$t_{ij}$. For the random chain the ratio of $\Gamma$ to the
spread of tunneling couplings turns out to be crucial. Clearly,
if the spread in tunneling couplings is much smaller than
$\Gamma$, the average conductance resembles that of the
homogenous infinite chain, which we have verified by numerical
simulation.
\begin{figure}[ptb]
\centering
\includegraphics[width=.5\textwidth]{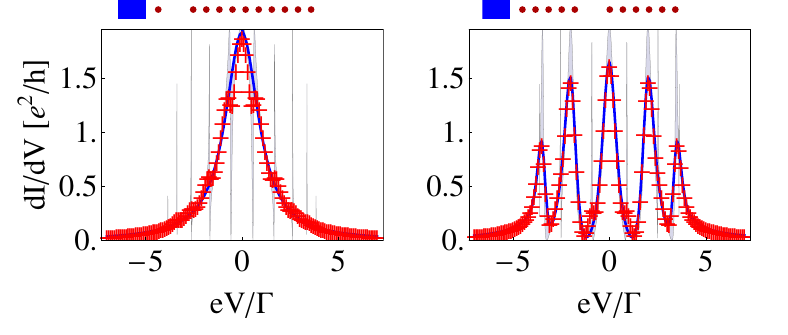} \caption{ (Color online)
Both panels show the conductance
for a chain with 10 sites and one weak link with tunnel coupling
$t_\mathrm{weak}=0.25\Gamma$, while the other links have $t=\Gamma$.
The weak links are the 1-2 and 5-6 connections for the left/right panel, respectively. The thick (blue) curve is
the conductance for $kT=0.1\Gamma$, the thin gray curve for $kT=0$, while the crosses show the conductance
for the chain truncate at the weak link also with $kT=0.1\Gamma$. It is clearly seen that the
truncated approximation works well, because the chain after the truncations leads to structure
barely resolvable, because $k_BT$ is not much smaller than the width ($\approx 0.125$) of the additional resonances.}
\label{fig:gweak}
\end{figure}

In contrast, with large fluctuations in the tunneling
couplings, \textit{the infinite chain will be effectively
truncated into a finite chain, where one of the tunneling
coupling happens to be much smaller than $k_BT$}. To see this,
invert the matrix in Eq.~\eqref{GR} and pull out the dependence
on the weak link and write is as
\begin{equation}\label{g11det}
    G^R(11,\omega)=\frac{2D_{2,\infty}}{D_{1,\infty}-4t_{n,n+1}^2
    D_{1,n}D_{n+1,\infty} +i\Gamma D_{2,\infty}},
\end{equation}
where $t_\mathrm{weak}=t_{n,n+1}$ is the weak link, and where
$D_{i,j}$ is the determinant of the matrix
$(\omega-2\mathbf{t})$ for the isolated chain between sites $i$
and $j$, but with $t_{n,n+1}=0$. The weak link has two effects,
1) it gives small shifts of the existing resonances and 2) it
creates new resonances. The new resonances, however, have
widths that scale with the square of the weak coupling
$\Gamma_\mathrm{weak}\propto t_\mathrm{weak}^2/\langle t\rangle
\Gamma$, where $\langle t\rangle$ denotes typical couplings in
the first part of the chain. Therefore the new resonances
introduced by the chain after the weak link is not resolved if
$k_Bt\gtrsim\Gamma_\mathrm{weak}$. An example of this is shown
in Fig.~\ref{fig:gweak}, where the conductance of a truncated
chain is compared with that of a full chain.

In summary, the differential conductance for a junction between
a normal metal and topological superconductor hosting a network
of Majorana bound states has been studied. Different
configurations of the interacting network of bound states give
rise to distinct tunneling spectra. Long chains with
fluctuating tunneling couplings is truncated into a finite
chain once a coupling becomes smaller than a certain critical
value, determined by temperature.

\acknowledgments

C.M. Marcus and  M. Leijnse are gratefully acknowledged for
stimulating discussions. The work was supported by The Danish
Council for Independent Research $|$ Natural Sciences.

%\bibliography{../Majorana}

\begin{thebibliography}{10}

\bibitem{Nayak2008} C. Nayak {\it et~al.}, Rev.~Mod.~Phys. {\bf
    80},  1083  (2008).

\bibitem{Kitaev2001} A.~Y. Kitaev, Physics-Uspekhi {\bf 44},
    131  (2001).

\bibitem{Fu2008} L. Fu and C. L. Kane, Physical Review Letters
    {\bf 100},  096407  (2008).

\bibitem{Oreg2010} Y. Oreg, G. Refael, and F.~V. Oppen, Arxiv
    preprint arXiv:1003.1145  (2010).

\bibitem{alicea2010} J. Alicea, Phys.~Rev.~B {\bf 81},  125318
    (2010).

\bibitem{Lutchyn2010} R.~M. Lutchyn, J.~D. Sau, and S.~Das Sarma, Phys.~Rev.~Lett. {\bf 105},  077001
  (2010).

\bibitem{Sau2010} J. Sau {\it et~al.}, Arxiv preprint arXiv:
    {\bf 26506},    (2010).

\bibitem{Fu2009} L. Fu and C. L. Kane, Phys.~Rev.~Lett. {\bf
    102},
    216403  (2009).

\bibitem{Akhmerov2009} A.~R. Akhmerov, J. Nilsson, and C.~W.~J.
    Beenakker, Phys.~Rev.~Lett. {\bf 102},
   216404  (2009).

\bibitem{Fu2010} L. Fu, Phys.~Rev.~Lett. {\bf 104},  056402
    (2010).

\bibitem{Law2009} K. T. Law, P. A. Lee, and T. K. Ng, Phys.~Rev.~Lett.
    {\bf 103},  237001  (2009).

\bibitem{Linder2010} J. Linder, Y. Tanaka, T. Yokoyama, A. Sudb{\o}, and N. Nagaosa, Phys.~Rev.~Lett.
    {\bf 104},  067001  (2010).

\bibitem{Bolech2007} C. J. Bolech and E. Demler, Phys.~Rev.~Lett.
    {\bf 98},  237002  (2007).

\bibitem{Nilsson2008} J. Nilsson, A.~R. Akhmerov, and C.~W.~J.
    Beenakker, Phys.~Rev.~Lett. {\bf 101},
   120403  (2008).

\bibitem{Shivamoggi2010} V. Shivamoggi, G. Refael, and J. E. Moore, Phys.~Rev.~B {\bf 82},  041405  (2010).

\bibitem{BTK} G.~E. Blonder, M. Tinkham, and T.~M. Klapwijk,
    Phys.~Rev.~B {\bf 25},  4515
  (1982).

\bibitem{multi} by generalizing the $k$-sum in
    Eq.~\eqref{Gammadef} to run over the different
  contacts as well and inserting a matrix that specifies the measurement
  contact in the trace in Eq.~\eqref{Mdef}.

\bibitem{branchcuts} For $|\omega|<4t$ this is
  $\tilde{g}_{22}=(\omega-i\sqrt{(4t)^2-\omega^2})/8t^2$, whereas for
  $|\omega|>4t$ the result is
  $\tilde{g}_{22}=(\omega-\mathrm{sign}(\omega)\sqrt{\omega^2-(4t)^2})/8t^2$.

\bibitem{levelcrossing} I. Blake and W. Lindsey, IEEE
    Transactions on Information Theory \textbf{19},
  295 (1973).
\end{thebibliography}
%\bibliographystyle{../prsty}

\appendix

\section{Current formula for a Majorana state network coupled to
normal-metal electrodes}

The low energy model Hamiltonian projected onto the Majorana
subspace is
\begin{subequations}
\label{H}
\begin{align}
H  &  =H_{N}+H_{S}+H_{T}\label{Hteff}\\
H_{N}  &  =\sum_{k\sigma}\varepsilon_{p}^{{}}c_{p}^{\dagger}c_{p}^{{}},\\
H_{S}  &  =\frac{i}{2}\sum_{ij}t_{ij}\gamma_{i}^{{}}\gamma_{j}^{{}},\\
H_{T}  &  =\sum_{pi}(V_{pi}^{\ast}c_{p}^{\dagger}-V_{pi}^{{}}c_{p}^{{}}
)\gamma_{i},
\end{align}
where $p=k,\sigma.$ In the non-equilibrium case the voltage on
the normal
metal is $eV$. The coupling matrix $t_{ij}$ is real and obeys $t_{ij}=-t_{ji}
$. The current operator is
\end{subequations}
\begin{align}
I  &  =-e\dot{N}=-ie[H_{T},N]/\hbar\nonumber\\
&  =\frac{ie}{\hbar}\sum_{pi}\left(  V_{pi}^{\ast}c_{p}^{\dagger}\gamma
_{i}-V_{pi}^{{}}\gamma_{i}c_{p}^{{}}\right) \nonumber\\
&  =\frac{e}{\hbar}\sum_{pi}\left(  V_{pi}^{\ast}G_{ip}
^{<}(0,0)-V_{pi}^{{}}G_{pi}^{<}(0,0)\right),
\end{align}
and
\begin{subequations}
\begin{align}
G_{ip}^{<}(t,t^{\prime})  &  =i\left\langle c_{p}^{\dagger}(t^{\prime}
)\gamma_{i}(t)\right\rangle ,\\
G_{pi}^{<}(t,t^{\prime})  &  =i\left\langle \gamma_{i}(t)c_{p}(t^{\prime
})\right\rangle .
\end{align}
The corresponding Keldysh contour Green's functions are
\end{subequations}
\begin{subequations}
\begin{align}
G_{ip}^{{}}(\tau,\tau^{\prime})  &  =-i\left\langle T_{K}^{{}}(\gamma_{i}
(\tau)c_{p}^{\dagger}(\tau^{\prime}))\right\rangle ,\\
G_{pi}^{{}}(\tau,\tau^{\prime})  &  =-i\left\langle T_{K}^{{}}(c_{p}^{{}}
(\tau)\gamma_{i}(\tau^{\prime}))\right\rangle .
\end{align}
Below we will also need
\[
\bar{G}_{pi}^{{}}(\tau,\tau^{\prime})=-i\left\langle T_{K}^{{}}(c_{p}
^{\dagger}(\tau)\gamma_{i}(\tau^{\prime}))\right\rangle ,
\]
the Majorana Green's function
\begin{equation}
G_{ij}(\tau,\tau^{\prime})=-i\left\langle T_{K}^{{}}(\gamma_{i}^{{}}
(\tau)\gamma_{j}^{{}}(\tau^{\prime}))\right\rangle ,
\end{equation}
and the free lead electron Green's functions
\begin{align}
G_{p}^{0}(\tau,\tau^{\prime})  &  =-i\left\langle T_{K}^{{}}(c_{p}^{{}}
(\tau)c_{p}^{\dagger}(\tau^{\prime}))\right\rangle _{0},\\
\bar{G}_{p}^{(0)}(\tau,\tau^{\prime})  &  =-i\left\langle T_{K}\left(
c_{p}^{\dagger}(\tau)c_{p}^{{}}(\tau^{\prime})\right)  \right\rangle _{0},
\end{align}
where the unperturbed expectation value $\left\langle
\cdot\right\rangle _{0}$ is for the situation with $H_{T}=0$.
The self energies
\end{subequations}
\begin{subequations}
\begin{align}
\Sigma_{ij}^{e}(\tau,\tau^{\prime})  &  =\sum_{p}V_{pi}^{{}}V_{pj}^{\ast
}G_{p}^{0}(\tau,\tau^{\prime}),\\
\Sigma_{ij}^{h}(\tau,\tau^{\prime})  &  =\sum_{p}V_{pi}^{\ast}V_{pj}^{{}}
\bar{G}_{p}^{(0)}(\tau,\tau^{\prime}),
\end{align}
will also appear.

\begin{widetext}
Now return to the mixed Green's functions $G_{ip},G_{pi}$, and
$\bar{G}_{pi}$. From diagrammatics, direct expansion, or
equation of motion, we obtain
\begin{subequations}
\begin{align}
\sum_{p^{\prime}i}G_{ip}^{{}}(\tau,\tau^{\prime})V_{pi}^{\ast} &  =\sum
_{ij}\int d\tau^{\prime\prime}~G_{ij}(\tau,\tau^{\prime\prime})\Sigma_{ji}
^{e}(\tau^{\prime\prime},\tau^{\prime})=\int d\tau^{\prime\prime
}~\mathrm{Tr}\left[  \mathbf{G}(\tau,\tau^{\prime\prime})\boldsymbol{\Sigma}
^{e}(\tau^{\prime\prime},\tau^{\prime})\right]  ,\\
\sum_{p^{\prime}i}V_{pi}G_{pi}^{{}}(\tau,\tau^{\prime}) &  =\sum_{ij}\int
d\tau^{\prime\prime}~\Sigma_{ij}^{e}(\tau,\tau^{\prime\prime})G_{ji}%
(\tau^{\prime\prime},\tau^{\prime})=\int d\tau^{\prime\prime}~\mathrm{Tr}%
\left[  \boldsymbol{\Sigma}^{e}(\tau,\tau^{\prime\prime})\mathbf{G}(\tau^{\prime\prime
},\tau^{\prime})\right],     \\
\sum_{p^{\prime}i}V_{pi}^*\bar{G}_{pi}^{{}}(\tau,\tau^{\prime}) &  =-\sum_{ij}\int
d\tau^{\prime\prime}~\Sigma_{ij}^{h}(\tau,\tau^{\prime\prime})G_{ji}%
(\tau^{\prime\prime},\tau^{\prime})=-\int d\tau^{\prime\prime}~\mathrm{Tr}%
\left[  \boldsymbol{\Sigma}^{h}(\tau,\tau^{\prime\prime})\mathbf{G}(\tau^{\prime\prime
},\tau^{\prime})\right] .
\end{align}
\label{SG}
\end{subequations}
\end{widetext}

With this the current becomes
\end{subequations}
\[
I=\frac{e}{\hbar}\int\frac{d\omega}{2\pi}\mathrm{Tr}\left[  \left(
\mathbf{G}\boldsymbol{\Sigma}^{e}-\boldsymbol{\Sigma}^{e}\mathbf{G}\right)
_{\omega}^{<}\right]  .
\]
Next, let us find the Dyson equation for the Majorana Green's
function. Its equation of motion is
\begin{equation}
i\partial_{\tau}G_{ij}(\tau,\tau^{\prime})=2\delta_{ij}\delta(\tau
,\tau^{\prime})+i\left\langle T_{K}^{{}}([H,\gamma_{i}^{{}}](\tau)\gamma
_{j}^{{}}(\tau^{\prime}))\right\rangle .
\end{equation}
\begin{widetext}
The factor of 2 because
$\{\gamma_{i},\gamma_{j}\}=2\delta_{ij}$. The commutators:
\begin{align}
\lbrack H_{0},\gamma_{i}]  &  =\frac{i}{2}\sum_{i^{\prime}j^{\prime}%
}t_{i^{\prime}j^{\prime}}[\gamma_{i^{\prime}}^{{}}\gamma_{j^{\prime}}^{{}%
},\gamma_{i}]
=i\sum_{j}\left(  t_{ji}-t_{ij}\right)  \gamma_{j}^{{}}
=-2i\sum_{j}t_{ij}\gamma_{j}^{{}}.\\
[H_{T},\gamma_{i}] & =\sum_{pj}\left[  (V_{pj}^{\ast}c_{p}^{\dagger}%
-V_{pj}^{{}}c_{p}^{{}})\gamma_{j},\gamma_{i}\right]
=2\sum_{p}(V_{pi}^{\ast}c_{p}^{\dagger}-V_{pi}^{{}}c_{p}^{{}}).
\end{align}

Again the factors of 2 come from the unusual commutation
relation. With this the equation of motion becomes
\begin{equation}
i\partial_{\tau}G_{ij}(\tau,\tau^{\prime})=2\delta_{ij}\delta(\tau
,\tau^{\prime})+2i\sum_{j^{\prime}}t_{ij^{\prime}}G_{j^{\prime}j}(\tau
,\tau^{\prime})+2\sum_{p}\left(  V_{pi}^{{}}G_{pj}(\tau,\tau^{\prime}%
)-V_{pi}^{\ast}\bar{G}_{pj}(\tau,\tau^{\prime})\right).
\end{equation}
By Eq.~\eqref{SG} the Dyson equation closes
\begin{equation}
i\partial_{\tau}G_{ij}(\tau,\tau^{\prime})=2\delta_{ij}\delta(\tau
,\tau^{\prime})+2i\sum_{j^{\prime}}t_{ij^{\prime}}G_{j^{\prime}j}(\tau
,\tau^{\prime})+2\sum_{j'}\int d\tau^{\prime\prime}\left( \Sigma^e_{ij\prime}(\tau,\tau^{\prime\prime})
+\Sigma^h_{ij\prime}(\tau,\tau^{\prime\prime})\right)
G_{j\prime j}(\tau^{\prime\prime},\tau^{\prime}),
\end{equation}
\end{widetext}
or in shorthand notation
\begin{equation}
\left(  i\partial_{\tau}-2i\mathbf{t}-2\boldsymbol{\Sigma}\right)  G=2,
\end{equation}
which has the solution
\begin{equation}
\label{Gsol}\mathbf{G}=\mathbf{G}^{0}+\mathbf{G}^{0}\boldsymbol{\Sigma
}\mathbf{G},
\end{equation}
where the unperturbed Majorana Green's function is
\begin{equation}
\left(  i\partial_{\tau}-2i\mathbf{t}\right)\mathbf{G}^{0}=2,
\end{equation}
and where the Majorana self energy is
\begin{equation}
\label{Mself}\boldsymbol{\Sigma}=\boldsymbol{\Sigma}^{e}+\boldsymbol{\Sigma}^{h}.
\end{equation}
The same result can be derived using diagrammatics, in which
case the factors of 2 then comes from the 2 ways 4 Majoranas
can pair.

The retarded components of the self energy are
\begin{align}
\Sigma_{ij}^{eR}\left(  \omega\right)   &  =\sum_{p}\frac{V_{pi}^{{}}%
V_{pj}^{\ast}}{\omega\!-\!\varepsilon_{p}\!+\!i\eta}=i\frac{\Gamma_{ij}%
(\omega)}{2}+\Lambda_{ij}(\omega),\\
\Sigma_{ij}^{hR}\left(  \omega\right)   &  =\sum_{p}\frac{V_{pi}^{\ast}%
V_{pj}^{{}}}{\omega\!+\!\varepsilon_{p}\!+\!i\eta}=i\frac{\Gamma_{ji}%
(-\omega)}{2}-\Lambda_{ji}(-\omega),
\end{align}
and as usual
$\boldsymbol{\Sigma}^{A}=(\boldsymbol{\Sigma}^{R})^{\dagger}$.
Here the $\mathbf{\Gamma}$ and $\mathbf{\Lambda}$ matrices are
defined as
\begin{align}
\Gamma_{ij}(\omega) &  =2\pi\sum_{p}V_{pi}^{{}}V_{pj}^{\ast}\delta\left(
\omega-\varepsilon_{p}\right)  ,\\
\Lambda_{ij}(\omega) &  =\mathcal{P}\!\int\frac{d\omega^{\prime}}{2\pi}%
\frac{\Gamma_{ij}(\omega^{\prime})}{\omega-\omega^{\prime}},
\end{align}
and they are both Hermitian matrices, so that $\boldsymbol{\Sigma}%
^{R}-\boldsymbol{\Sigma}^{A}=i\left(
\mathbf{\Gamma}_{\omega}+\mathbf{\Gamma
}_{-\omega}^{\ast}\right)  $ .

The lesser components are
\begin{align}
\Sigma_{ij}^{e<}(\omega) &  =i\Gamma_{ij}^{{}}\left(  \omega\right)
f(\omega-eV),\\
\Sigma_{ij}^{h<}(\omega) &  =i\Gamma_{ij}^{\ast}\left(  -\omega\right)
(1-f(\omega-eV)).
\end{align}

\newpage
Now go back to current and use that
$(BC)^{<}=B^{R}C^{<}+B^{<}C^{A}$ to get
\begin{equation}
I=\frac{e}{\hbar}\int\frac{d\omega}{2\pi}~\mathrm{Tr}\left[  \left(
\mathbf{G}_{\omega}^{R}-\mathbf{G}_{\omega}^{A}\right)  \boldsymbol{\Sigma
}_{\omega}^{e<}+\mathbf{G}_{\omega}^{<}\left(  \boldsymbol{\Sigma}_{\omega
}^{eA}-\boldsymbol{\Sigma}_{\omega}^{eR}\right)  \right]
\end{equation}
In equilibrium $\mathbf{G}_{\omega}^{<,eq}=-i\left(  \mathbf{G}_{\omega}%
^{R}-\mathbf{G}_{\omega}^{A}\right)  f_{\omega}$ and the
current is zero. The Majorana lesser function is
\begin{equation}
\mathbf{G}_{\omega}^{<}=\mathbf{G}_{\omega}^{R}\Sigma_{\omega}^{<}%
\mathbf{G}_{\omega}^{A}=i\mathbf{G}_{\omega}^{R}\left(  \mathbf{\Gamma
}_{\omega}f_{\omega-eV}+\mathbf{\Gamma}_{-\omega}^{\ast}f_{-\omega+eV}\right)
\mathbf{G}_{\omega}^{A},
\end{equation}
since $f_{\omega-eV}=1-f_{-\omega+eV}.$ Together with
\begin{equation}
\mathbf{G}_{\omega}^{R}-\mathbf{G}_{\omega}^{A}=\mathbf{G}_{\omega}^{R}\left(
\mathbf{\Sigma}_{\omega}^{R}-\mathbf{\Sigma}_{\omega}^{A}\right)
\mathbf{G}_{\omega}^{A}=i\mathbf{G}_{\omega}^{R}\left(  \boldsymbol{\Gamma
}_{\omega}^{{}}+\boldsymbol{\Gamma}_{-\omega}^{\ast}\right)  \mathbf{G}%
_{\omega}^{A},
\end{equation}
this finally leads to%
\begin{equation}
I=\frac{e}{h}\int d\omega~M(\omega)(f_{-\omega+eV}-f_{\omega-eV}),
\end{equation}
with
\begin{equation}
M(\omega)=\mathrm{Tr}\left[  \mathbf{G}_{\omega}^{R}\mathbf{\Gamma}_{-\omega
}^{\ast}\mathbf{G}_{\omega}^{A}\boldsymbol{\Gamma}_{\omega}^{{}}\right]  ,
\end{equation}
which is the final general result for the current from a normal
metal into a network of Majorana fermions.

The retarded Green's function is
\begin{equation}
\mathbf{G}_{\omega}^{R}=2\left(  \omega-2i\mathbf{t}+i\left(  \mathbf{\Gamma
}_{\omega}+\mathbf{\Gamma}_{-\omega}^{\ast}\right)  -\left(  \mathbf{\Lambda
}_{\omega}-\mathbf{\Lambda}_{-\omega}^{\ast}\right)  \right)  ^{-1}.
\end{equation}
In the electron-hole symmetric case $\mathbf{\Gamma
}_{\omega}=\mathbf{\Gamma}_{-\omega}^{\ast}=\mathbf{\Gamma}_{-\omega}^{T}$,
and hence
$\mathbf{\Lambda}_{\omega}=-\mathbf{\Lambda}_{-\omega}^{\ast},$
it follows that
\begin{equation}
\mathbf{G}_{-\omega}^{R}=-\left(  \mathbf{G}_{\omega}^{R}\right)  ^{\ast
}=-\left(  \mathbf{G}_{\omega}^{A}\right)  ^{T},\label{GRGA}%
\end{equation}
and therefore
\begin{align}
M(-\omega)  & =\mathrm{Tr}\left[  \mathbf{G}_{-\omega}^{R}\mathbf{\Gamma
}_{\omega}^{\ast}\mathbf{G}_{-\omega}^{A}\boldsymbol{\Gamma}_{-\omega}\right]
=\mathrm{Tr}\left[  \mathbf{G}_{\omega}^{AT}\mathbf{\Gamma}_{-\omega}^{\ast
T}\mathbf{G}_{\omega}^{RT}\boldsymbol{\Gamma}_{\omega}^{T}\right]
\nonumber\\
& =\mathrm{Tr}\left[  (\boldsymbol{\Gamma}_{\omega}^{{}}\mathbf{G}_{\omega
}^{R}\mathbf{\Gamma}_{-\omega}^{\ast}\mathbf{G}_{\omega}^{A})^{T}\right]
=M(\omega).
\end{align}
Therefore, if the leads do not break electron-hole symmetry the
current is anti-symmetric $I(V)=-I(-V).$

\end{document}